\documentclass[10pt,aps,rmp,onecolumn,superscriptaddress,floatfix,nofootinbib,showkeys,notitlepage,preprintnumbers]{revtex4-1}
\usepackage{amsmath,amssymb,amsfonts}
\usepackage{graphicx,color}
\usepackage{multirow}
\usepackage{bm}
\newcommand{\ie}{\emph{i.e.}}
\newcommand{\eg}{\emph{e.g.}}

\def\Nck{{N_{\mathcal{C}}^{[k]}(t)}}
\def\Npk{{N_{\mathcal{P}}^{[k]}(t)}}
\def\Ncks{{N_{\mathcal{C}}^{[k]}}}
\def\Npks{{N_{\mathcal{P}}^{[k]}}}
\def\rhocck{{\rho_{\mathcal{CC}}^{[k]}(t)}}
\def\rhocpk{{\rho_{\mathcal{CP}}^{[k]}(t)}}
\def\rhoppk{{\rho_{\mathcal{PP}}^{[k]}(t)}}
\def\rhoccks{{\rho_{\mathcal{CC}}^{[k]}}}
\def\rhocpks{{\rho_{\mathcal{CP}}^{[k]}}}
\def\rhoppks{{\rho_{\mathcal{PP}}^{[k]}}}
\def\rhocc1{{\rho_{\mathcal{CC}}^{[1]}(t)}}
\def\rhocp1{{\rho_{\mathcal{CP}}^{[1]}(t)}}
\def\rhopp1{{\rho_{\mathcal{PP}}^{[1]}(t)}}
\def\ER{Erd\H{o}s-R\'{e}nyi\ }

\begin{document}
\title{Structural changes in the interbank market across the financial crisis from multiple core-periphery analysis}

\author{Sadamori Kojaku} 
\affiliation{\small Department of Engineering Mathematics University of Bristol - Bristol BS81UB (United Kingdom)}
\affiliation{\small Core Research for Evolutional Science and Technology (CREST) JST - Kawaguchi-shi Saitama 332-0012 (Japan)}
\author{Giulio Cimini}\email{giulio.cimini@imtlucca.it}
\affiliation{\small IMT School for Advanced Studies - 55100 Lucca (Italy)}
\affiliation{\small Istituto dei Sistemi Complessi (ISC)-CNR - 00185 Rome (Italy)}
\author{Guido Caldarelli}
\affiliation{\small IMT School for Advanced Studies - 55100 Lucca (Italy)}
\affiliation{\small Istituto dei Sistemi Complessi (ISC)-CNR - 00185 Rome (Italy)}
\affiliation{\small London Institute for Mathematical Sciences - London W1K2XF (United Kingdom)}
\affiliation{\small European Centre for Living Technologies - 30124 Venice  (Italy)}
\author{Naoki Masuda}
\affiliation{\small Department of Engineering Mathematics University of Bristol - Bristol BS81UB (United Kingdom)}

\begin{abstract}
Interbank markets are often characterised in terms of a core-periphery network structure, with a highly interconnected core of banks holding the market together, 
and a periphery of banks connected mostly to the core but not internally. This paradigm has recently been challenged for short time scales, 
where interbank markets seem better characterised by a bipartite structure with more core-periphery connections than inside the core. 
Using a novel core-periphery detection method on the eMID interbank market, we enrich this picture by showing that the network is actually characterised by multiple core-periphery pairs. 
Moreover, a transition from core-periphery to bipartite structures occurs by shortening the temporal scale of data aggregation. 
We further show how the global financial crisis transformed the market, in terms of composition, multiplicity and internal organisation of core-periphery pairs. 
By unveiling such a fine-grained organisation and transformation of the interbank market, 
our method can find important applications in the understanding of how distress can propagate over financial networks.
\end{abstract}

\maketitle
\vspace{-1cm}
\subsubsection*{Key messages}
\begin{itemize}
\item We study the eMID interbank market using a novel core-periphery detection method to unveil its high-order network organization.
\item We reveal multiple core-periphery pairs, and detect both tiered and bow-tie (bipartite) structures.
\item We reveal a topological transition of the market at the global financial crisis, in terms of network segregation and organization.
\end{itemize}

\section{Introduction}

The financial turmoils of the last decade highlighted the inherent fragility of the financial system arising from the complex interconnections between financial institutions 
(see, \eg, \citet{Gai2011,Glasserman2015,Battiston2016complexity}). 
Several contributions have shown that systemic risk can result both from direct exposures to bilateral contracts and from indirect exposures to common assets 
\citep{Kaufman1994,Allen2000,Furfine2003,Cifuentes2005,Elsinger2006,Lau2009,Gai2010,Haldane2011,Battiston2012,Cont2013,Georg2013,Caccioli2014,Acemoglu2015,Bardoscia2015,Greenwood2015,Amini2016,Cimini2016,Cont2016,Gualdi2016,Bardoscia2017}. 
Much research has thus been carried out by academics and regulators to characterise the emerging network structure of financial markets \citep{Boss2004,Iori2006,Nier2007,Cocco2009}.
A parallel line of research consisted in developing methods to reconstruct the network structure from available data---usually, 
the balance sheet composition of financial institutions, since data on individual exposures are often privacy-protected 
(see \citet{Anand2015,Cimini2015systemic,Cimini2015estimating,DiGiangi2015,Gandy2015X,Squartini2017ECAPM} among the most recent contributions, 
and \citet{Anand2017} for a comparison of most of these methods).

In this context, much attention has been devoted to studying the interbank lending market, namely the network of financial interconnections between banks resulting from unsecured overnight loans 
\citep{Rochet1996,Freixas2000} (see \citet{Huser2015} for a recent overview of the field). Within this market, banks temporarily short on liquidity borrow money for a specified term 
from other banks having excess liquidity, that in turn receive an interest on the loan. Thus this market plays a crucial role by allowing banks cope with liquidity fluctuations \citep{Allen2014}. 
However, the interbank market is rather sensitive to market movements, 
and it can dry up under exceptional circumstances \citep{Brunnermeier2009a}---becoming a main vehicle for distress propagation within the financial system \citep{Angelini2011,Acharya2013,Berrospide2013}. 
Indeed, during the 2007/2008 global financial crisis, worries of counterparty creditworthiness and fire sales spillovers led banks to hoard liquidity, 
causing the freeze of credit to the financial system and the real economy \citep{Diamond2009,Adrian2010,Krause2012,Gale2013,Serri2017}.

Because of its structure, the interbank market can be represented as a network, where interbank loans constitute the direct exposures between banks 
\citep{DeMasi2006,Leon2014,Iori2015}. \citet{Finger2013} analysed the network properties of the electronic market for interbank deposits (eMID) at various temporal scales of data aggregation, 
showing that the network appears random at the daily level, but contains significant non-random structures for longer aggregation periods. 
Several empirical studies further reported that interbank markets have a {\em core-periphery} (or tiered) structure, featuring a core of highly interconnected banks 
and a periphery of banks connected mostly to the core but not to other peripheral banks \citep{Iori2008,Bech2010,Soramaki2007,Craig2014,Veld2014,Langfield2014,MartinezJaramillo2014,Fricke2015,Silva2016}. 
\citet{Craig2014} argued that tiering derives from core banks acting as intermediaries between periphery banks and thus holding together the interbank market. 
Unfortunately, according to \citet{Lee2013} core-periphery structures with a deficit money centre bank do give rise to the highest level of systemic liquidity shortage. 
In fact \citet{Fricke2015} showed that, during the global financial crisis, the reduction in interbank market size was mainly due to core banks reducing their lending. 
\citet{Verma2016} explained the emergence of core-periphery using an economic argument based on trade-offs between the profit of establishing connections and the cost for maintaining them. 

Recently, \citet{Barucca2016,Barucca2018} used a stochastic block-modelling approach to show that the core-periphery structure of eMID emerges only when data are aggregated for more than a week. 
They also showed that, for shorter aggregation periods, the network is instead better characterised by a {\em bipartite} (or bow-tie) structure, 
with connections established mainly between core and periphery banks. This means that for short time scales banks in eMID either lend or borrow---a rather different situation 
than a market where core banks intermediate between peripheral banks. 
Bipartitivity emerges even at longer aggregation periods when using an extended stochastic block model that takes into account the different numbers of connections that different banks have. 
\citet{Carreno2017} also used stochastic block models to analyse the interbank market of term deposits and derivatives in Chile at the daily level. 
Their approach permits to identify multiple cores and peripheries---though they rarely observed a second core---and allows to characterise different forms of interaction 
between core and periphery. Remarkably, they often found a core mostly funding itself from the periphery, rather than intermediating among peripheral banks.

In this work we add to the current discussion by employing a novel methodology able to reveal multiple core-periphery pairs in a network, automatically determining their number and size \citep{Kojaku2017}. 
Focusing on eMID, we show that the market actually features a main core-periphery pair mostly composed of Italian banks, 
a second smaller core-periphery pair of foreign banks, and many other smaller core-periphery pairs. 
The method also detects bipartite patterns for long (\ie, quarterly and monthly) as well as short (\ie, weekly and daily) data aggregation periods, which is consistent with what \citet{Barucca2016} found. 
By analysing the temporal evolution of the detected multiple core-periphery structures, we observe that the global financial crisis 
caused a considerable change in the core-periphery patterns of the network, for all the quantities we observed. 
Note that we compare these results with those found by well-known core-periphery detection methods: the Borgatti-Everett algorithm \citep{Borgatti2000} and MINRES \citep{Boyd2010,Lip2011}. 
These methods turn out to fail in detecting both core-periphery multiplicity and bipartitivity, as well as many structural changes of the network due to the crisis.

\section{Methods}

\subsection{eMID data}\label{sec:data}

The electronic Market for Interbank Deposits (eMID) is the first electronic market for interbank deposits established in the Euro area. 
Founded in Italy in 1990 for Italian Lira transactions, it was denominated in Euros in 1999, and to date currencies traded are Euros, USD 
and GBP.\footnote{A collateralised segment of e-MID (New MIC) was introduced in February 2009.} 
The market is subject to the supervision of the Bank of Italy, and is open to both Italian and foreign banks. 
eMID covers the whole domestic liquidity deposit market in Italy, as well as a significant share of the entire liquidity deposit market in the Euro area \citep{Beaupain2008}. 

Our dataset contains all interbank transactions finalised on eMID from January 1999 to September 2012. 
Each transaction record contains the IDs of the lender and the borrower banks, the amount of credit transferred and the time of the transaction.
These interbank loans constitute the bilateral exposures between banks, and the system can be conveniently described as a network.

Let $N$ be the number of banks in the system. We fix a time period of length $t$, and define the network at $t$ through the $N\times N$ {\em binary adjacency matrix} ${\bm A}(t)=\{A_{ij}(t)\}_{i,j=1}^N$. 
We set $A_{ij}(t)=1$ if banks $i$ and $j$ perform a trade on eMID at least once during $t$ (and say that these banks are {\em adjacent}), and $A_{ij}(t)=0$ otherwise. 
To ease the analysis, we ignore edge directions, so that the network matrices are symmetrical. 
We regard a bank as active at $t$ if it established at least one connection during the corresponding time period, and inactive otherwise. 
We construct quarterly, monthly, weekly and daily networks by setting the length of the time period to a quarter, a month, a week and a day, respectively.
We also construct a ``static'' network by setting $t$ equal to the entire time span of the dataset.

\subsection{Core-periphery structure}\label{sec:CPstruct}

\begin{figure}
\includegraphics[width=\hsize]{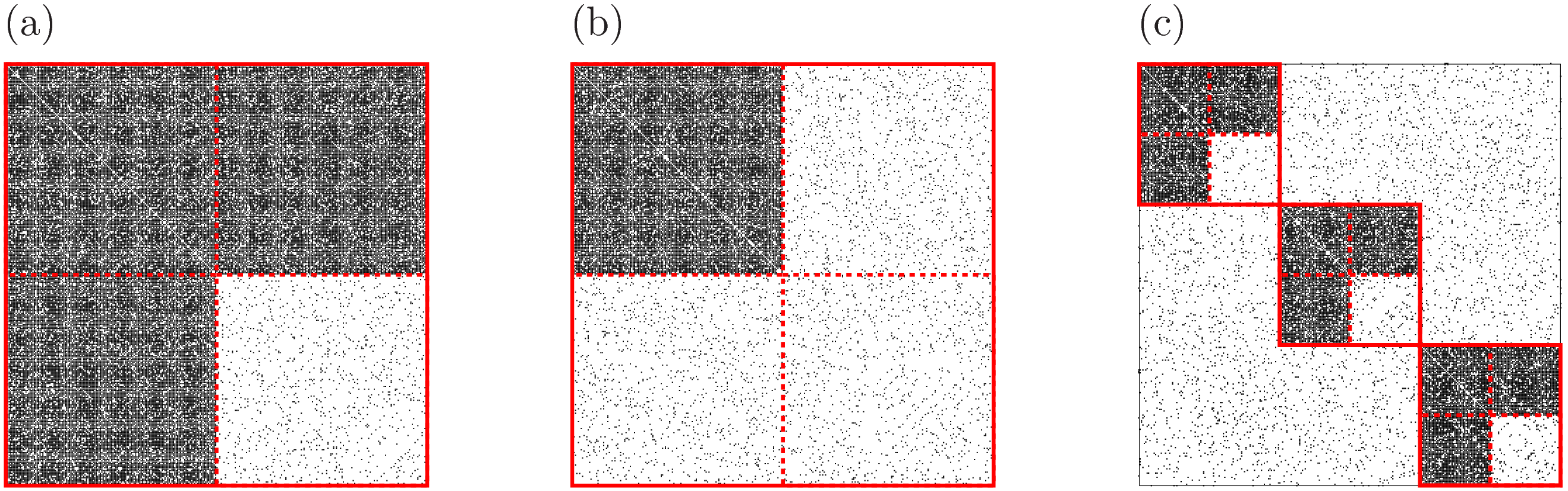}
\caption{Schematic illustration of three types of core-periphery structure. 
The filled (black) and empty (blank) cells indicate the presence and absence of connections, respectively. 
The solid squares delimit core-periphery pairs, and the dashed lines mark the boundaries between core and periphery.}
\label{fig:cp}
\end{figure}

As outlined in Figure~\ref{fig:cp}, a core-periphery structure based on the density of connections is composed of at least a pair of core and periphery blocks. 
The core block (in the upper left corner of Figs.~\ref{fig:cp}(a) and (b)) has many intra-block connections, whereas, the peripheral block (in the bottom right corner) has relatively few intra-block connections. 
Connections between core and periphery blocks (in the upper right and bottom left corners) may be abundant (Fig.~\ref{fig:cp}(a)) or not (Fig.~\ref{fig:cp}(b)). 
Networks may consist of a single core-periphery pair (Figs.~\ref{fig:cp}(a) and (b)) or of multiple core-periphery pairs (Fig.~\ref{fig:cp}(c)).

\subsection{Core-periphery detection}\label{sec:CPalgo}

In order to detect core-periphery structures we use the Borgatti-Everett (BE) \citep{Borgatti2000}, MINRES \citep{Boyd2010,Lip2011} and the Kojaku-Masuda (KM--ER) \citep{Kojaku2017} algorithms. 
In the following descriptions of the algorithms, we omit the time label $t$ to simplify the notation, recalling that all methods take as input a given adjacency matrix ${\bm A}(t)$.

In the BE algorithm, one considers an idealised single core-periphery pair, in which each core node is adjacent to all core and peripheral nodes, 
and each peripheral node is not adjacent to any other peripheral nodes. The adjacency matrix for the idealised single core-periphery pair is given by 
${\bm B}^{\text{BE}} = \{B^{\text{BE}}_{ij}\}_{i,j=1}^N$, with 
\begin{align}
	B _{ij}^{\text{BE}} \equiv 
	\left\{
	\begin{array}{ll}
	1 & \mbox{($x_i=1$ or $x_j = 1$)}, \\
	0 & \mbox{($x_i=x_j = 0$)}, 
	\end{array}
	\right. \label{eq:scp}
\end{align}
where $x_i = 1$ or $0$ if node $i$ is a core node or a peripheral node, respectively, and $i \neq j$.
The BE algorithm seeks ${\bm x}=[x_1, x_2, \ldots, x_N]$ by maximising the similarity between ${\bm B}^{\text{BE}}$ and the adjacency matrix ${\bm A}$ of the given network. 
The MINRES algorithm also seeks ${\bm x}$ by maximising the similarity between ${\bm B}^{\text{BE}}$ and ${\bm A}$, however ignoring the contribution from connections between the core and the periphery. 
As such, MINRES allows the core and periphery blocks to be sparsely connected (Fig.~\ref{fig:cp}(b)).
Note that both BE and MINRES algorithms identify only one pair of core and periphery in a given network. 

The KM--ER algorithm instead allows multiple core-periphery pairs (Fig.~\ref{fig:cp}(c)).
KM--ER considers a network composed of $C$ non-overlapping idealised core-periphery pairs, in which there is no connection between different core-periphery pairs. 
The adjacency matrix for the idealised core-periphery structure is given by ${\bm B}^{\text{KM--ER}} = \{B^{\text{KM--ER}}_{ij}\}_{i,j=1}^N$, with
\begin{align}
	B_{ij}^{\text{KM--ER}} \equiv 
	\left\{
	\begin{array}{ll}
	\delta(c_i, c_j) & \mbox{($x_i=1$ or $x_j = 1$)}, \\
	0 & \mbox{($x_i=x_j = 0$)}, 
	\end{array}
	\right.
\end{align}
where $c_i$ is the index of the core-periphery pair to which node $i$ belongs, $\delta(\cdot,\cdot)$ is Kronecker's delta, and $i\neq j$.
The KM--ER algorithm seeks ${\bm x}$ and ${\bm c}=[c_1, c_2, \ldots, c_N]$ by maximising
\begin{align}
	Q^{\rm cp} &\equiv \sum_{i=1} ^N \sum_{j=1} ^{i-1} A_{ij}B^{\text{KM--ER}}_{ij} - \sum_{i=1} ^N \sum_{j=1} ^{i-1} \rho\left({\bm A}\right) B^{\text{KM--ER}}_{ij} \nonumber \\
		   &= \sum_{i=1} ^N \sum_{j=1} ^{i-1} \left(A_{ij} - \rho\left({\bm A}\right) \right)(x_i + x_j - x_i x_j)\delta (c_i, c_j), \label{eq:qcp}
\end{align}
where $\rho \left({\bm A}\right)$ is the overall density of the network:
\begin{align}
	\rho \left({\bm A}\right) \equiv \dfrac{\sum_{i=1}^N \sum_{j=1}^{i-1} A_{ij} }{ \frac{1}{2} N(N-1) }.  \label{eq:rho}
\end{align}
In eq.~\eqref{eq:qcp}, the term $\sum_{i=1} ^N \sum_{j=1} ^{i-1} A_{ij}B^{\text{KM--ER}}_{ij}$ accounts for the number of connections that appear in both ${\bm A}$ and ${\bm B}^{\text{KM--ER}}$, 
whereas, the term $\sum_{i=1} ^N \sum_{j=1} ^{i-1} \rho({\bm A}) B^{\text{KM--ER}}_{ij}$ represents the number of edges for the \ER random graph that has the same number of edges as the given network.
A large $Q^{\rm cp}$ value indicates that the given network has strong core-periphery structure as compared to the \ER random graph.

To maximise $Q^{\rm cp}$, the KM--ER algorithm proceeds as follows \citep{Kojaku2017}.
First, we assign each node to a different core, \ie, $x_i = 1$ and $c_i = i$ for $1 \leq i \leq N$.
Then, we reassign the label $c_i$ to each node $i$ in a random order as follows.
For node $i$, we tentatively assign it to the core of the core-periphery pair to which a neighbour $j$ belongs, \ie, we set $x_i = 1$ and $c_i = c_j$.
Then, we tentatively assign node $i$ to the periphery of the core-periphery pair, \ie, we set $x_i = 0$ and $c_i = c_j$.
We repeat this procedure for each neighbour $j$ of node $i$ and adopt the tentative label---$(c_j, 0)$ or $(c_j, 1)$---that yields the largest increment in $Q^{\rm cp}$. 
If no relabelling increases $Q^{\rm cp}$, no update of $(c_i, x_i)$ is done.
After sweeping all nodes, we end the relabelling procedure if we have not changed any label in the current sweep. 
Otherwise, we sweep all nodes again in a new random order and repeat the procedure. 
We run this algorithm ten times starting from the same initial labels (\ie, $x_i = 1$ and $c_i = i$) and adopt the result giving the largest $Q^{\rm cp}$ value.

\subsubsection{Significance of a core-periphery pair}

In order to estimate the significance of the detected core-periphery pairs, we use the statistical method described in \citet{Boyd2006,Kojaku2017}.
Suppose a single core-periphery pair, described by ${\bm x}=[x_1, x_2, \ldots, x_N]$, is detected by an algorithm---say BE. 
First, we compute the quality of the core-periphery pair as  
\begin{align}
    \label{eq:bgquality}
    q({\bm x}; {\bm A}) = 
    \frac{
        \sum_{i=1}^{N} \sum_{j=1}^{i-1} 
			\left[
				A _{ij} - \rho\left({\bm A}\right)
			\right]
			\left[
				B^{\rm BE}_{ij} - \rho\left({\bm B}^{\rm BE}\right)
			\right]
    }{
        \sqrt{
		\sum_{i=1}^{N}\sum_{j=1}^{i-1}
		\left[A _{ij}-\rho\left({\bm A}\right)\right]^2}
	\sqrt{
		\sum_{i=1}^{N}\sum_{j=1}^{i-1}
		\left[B^{\rm BE}_{ij} - \rho\left({\bm B}^{\rm BE}\right)\right]^2}
    }.
\end{align}
Second, we generate $10^3$ randomized \ER random networks, each with the same number of nodes and edges of the original network; 
for each randomised network $\tilde {\bm A}$, we detect a single core-periphery pair $\tilde {\bm x}$ using BE, and compute its quality $q(\tilde{\bm x}; \tilde {\bm A})$.
Finally, we deem that the original core-periphery pair is significant if it has a larger quality than a fraction $1-\alpha$ of the single core-periphery pairs detected in the randomised networks, 
where $\alpha \in [0,1]$ is the significance level. 
If instead there are multiple core-periphery pairs detected, we apply the aforementioned statistical test to each of them.
We suppress the false positives owing to multiple comparisons using the {\v{S}}id{\'{a}}k correction \citep{Sidak1967}, \ie,  
$\alpha = 1-(1-\alpha')^{1/C}$, where $\alpha' = 0.05$ is the targeted significance level.
In the following, we refer to banks belonging to insignificant core-periphery pairs as {\em residual} banks.

\subsubsection{Characterisation of a core-periphery pair}

We use the density of connections to quantitatively characterise each block of a core-periphery pair. 
For the $k$-th core-periphery pair, \ie, including nodes with $c=k$, we define $\rhoccks$, $\rhocpks$ and $\rhoppks$ as the fraction of edges within the core block, 
that between the core and periphery blocks, and that within the periphery block, respectively. These quantities are given by: 
\begin{align}
	\rhoccks&\equiv 
		\frac{
			\sum_{i=1}^N \sum_{j=1}^{i-1} A_{ij}x_ix _j \delta(c_i, k)\delta(c_j, k)	
		}{ 
			\tfrac{1}{2}\Ncks(\Ncks-1)
		}, \\ 
	\rhoppks&\equiv 
		\frac{
			\sum_{i=1}^N \sum_{j=1}^{i-1} A_{ij}(1-x_i)(1-x_j) \delta(c_i, k)\delta(c_j,k)	
		}{ 
			\tfrac{1}{2}\Ncks(\Npks-1)
		}, \\ 
	\rhocpks&\equiv 
		\frac{
			\sum_{i=1}^N \sum_{j=1}^{i-1} A_{ij}(x_i + x_j -2x_ix_j)  \delta(c_i, k)\delta(c_j,k)	
		}{ 
			\Ncks\Npks
		}, 
\end{align}
where $\Ncks = \sum_{i=1}^N x_i\delta(c_i,k)$ and $\Npks = \sum_{i=1}^N (1-x_i)\delta(c_i,k)$ are the number of nodes of the core and periphery in the $k$th core-periphery pair, respectively.

\section{Results}

\subsection{Static network}\label{sec:static}

\begin{figure}
	\centering
	\includegraphics[width=\hsize]{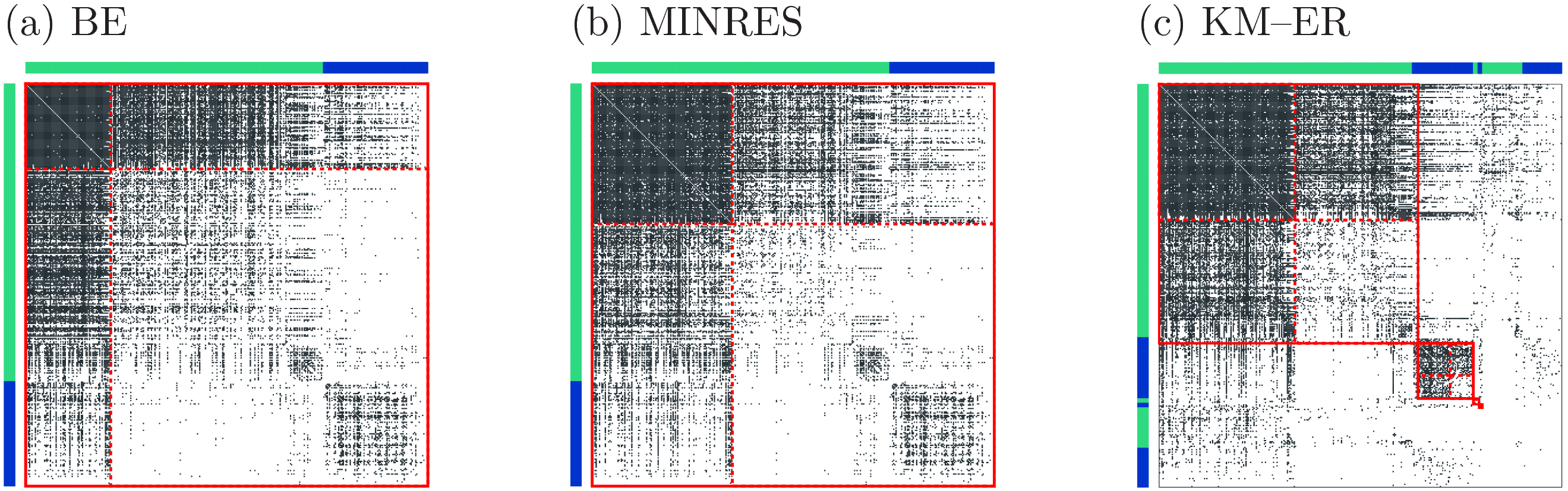}
\caption{Core-periphery pairs detected by the various algorithms. Colour labels on rows/columns indicates Italian (green) and foreign (blue) banks.}
\label{fig:adj}
\end{figure}

We start by analysing the static network, which is obtained by setting $t$ equal to the entire time span of the dataset.  
By definition, the BE and MINRES algorithms identify a single core-periphery pair (Figs.~\ref{fig:adj}(a) and \ref{fig:adj}(b)). In contrast, 
the KM--ER algorithm identifies four core-periphery pairs (Fig.~\ref{fig:adj}(c)). 
In this case, the largest and second largest core-periphery pairs mostly consist of Italian and foreign banks, respectively. 
Thus, the KM--ER algorithm discerns the network behaviour of Italian and foreign banks in an unsupervised way, which is consistent with previous studies \citep{Fricke2015}. 

Table~\ref{ta:prop} reports values of connection density for the core-periphery blocks detected by the three algorithms.
In all cases, except for star structure with only one core node, core-periphery pairs satisfy $\rhocck>\rhocpk>\rho(t)=0.28>\rhoppk$, where $\rho(t)$ is the overall density of the network. 
This means that the core-periphery characterisation of the network is rather marked for such a long aggregation period. 

\begin{table}
\centering
\caption{Properties of core-periphery pairs in the static network. The index $k$ labels the core-periphery pair.}
\label{ta:prop}
\begin{tabular}{lcccccc}
	& $k$ & $\Nck$ & $\Npk$ & $\rhocck$ & $\rhocpk$ & $\rhoppk$ \\ \hline \hline
{\bf BE}	& 1	& 74 & 276 & 0.98 & 0.50 & 0.11 \\ \hline
{\bf MINRES}	& 1	& 122 & 228 & 0.92 & 0.31 & 0.06 \\ \hline
\multirow{4}{*}{\bf KM--ER}	& 1	& 118 & 107 & 0.91 & 0.57 & 0.13 \\
	& 2	& 28 & 20 & 0.68 & 0.51 & 0.12 \\
	& 3	& 1 & 3 & 0 & 1 & 0 \\
	& 4	& 1 & 2 & 0 & 1 & 0 \\ \hline
\end{tabular}
\end{table}

\subsection{Temporal network}\label{sec:temporal}

\begin{figure}
\centering
\scalebox{0.8}{	\includegraphics[width=\hsize]{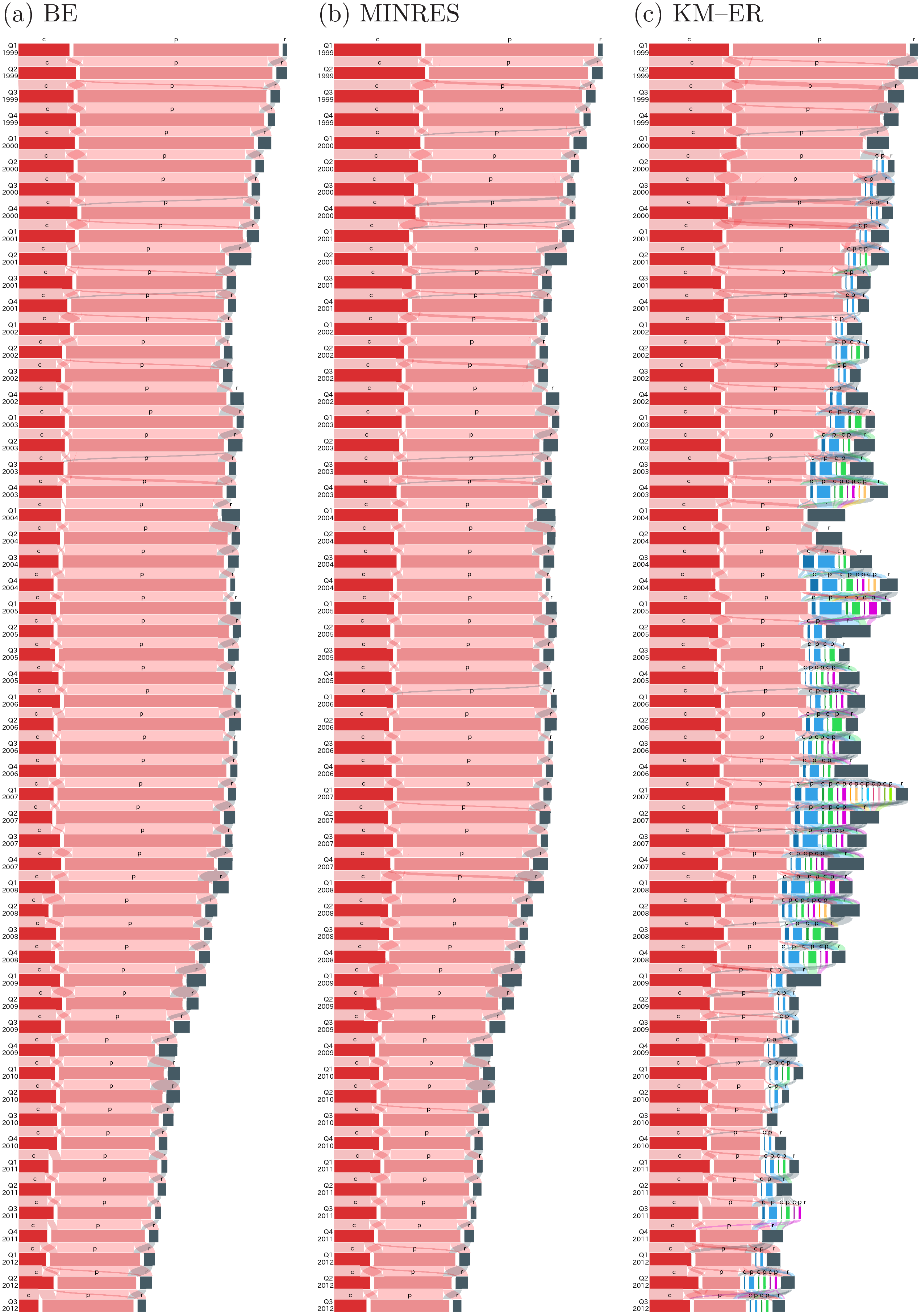} }
\caption{Core-periphery structures in quarterly temporal networks detected by the three algorithms.
Each rectangle indicates either a core, a periphery or the residual, as indicated by letters ``c'', ``p'' and ``r'', respectively.
Darker and lighter hue of the same colour indicate a core and the corresponding periphery, respectively.
Different colours (\eg, red, blue, green) indicate different core-periphery pairs, while grey marks the residual nodes.
The width of each rectangle represents the size (\ie, number of nodes) for each block.
The flow between the rectangles in consecutive time points (\ie, the shaded region) indicates the number of banks that move from one group to another.
Inactive banks at each quarter are not shown.}
\label{fig:allu_quarterly}
\end{figure}

\begin{figure}
\centering
\includegraphics[width=0.75\hsize]{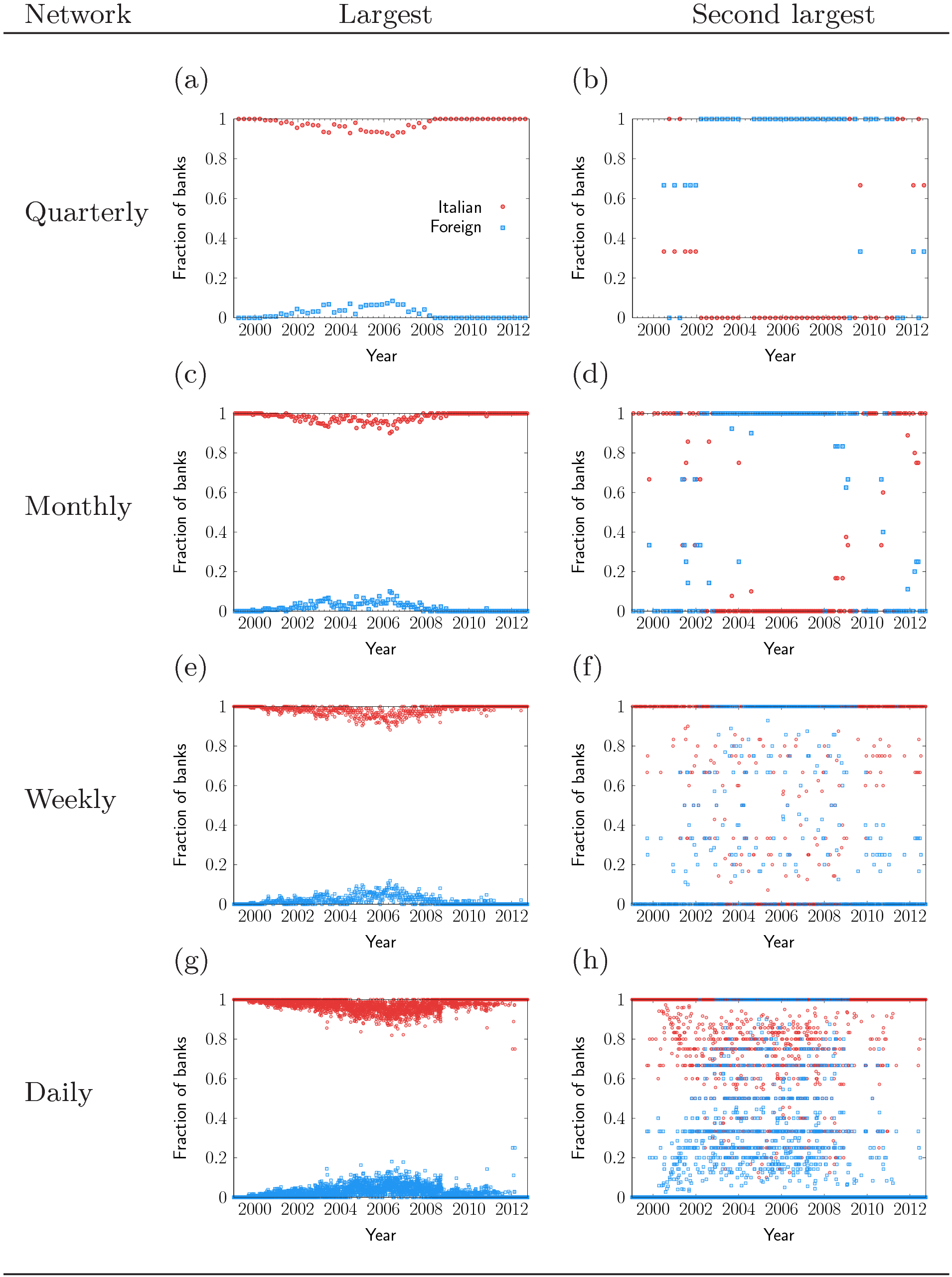}
\caption{Fraction of Italian and foreign banks in the largest and second largest core-periphery pairs detected by the KM--ER algorithm.}
\label{fig:italian_foreign_kmerb}
\end{figure}

We now move to the analysis of temporal networks, in which $t$ is set equal to either a quarter, a month, a week and a day. 
Figure~\ref{fig:allu_quarterly} shows how core-periphery structures detected by the three algorithms change over time, in the case of quarterly networks 
(refer to Figs.~\ref{fig:allu_beb_monthly}--\ref{fig:allu_kmerb_monthly} for the case of monthly networks).\footnote{Shorter time scales 
cannot be properly shown with these kinds of plots, but their properties will be as well analysed in the following.} 
We see that the KM--ER algorithm detects, almost always, a significantly large core-periphery pair (which we will refer to as {\em main} core-periphery pair), 
accompanied by a few small core-periphery pairs. These small pairs are very short-lived, as they persist for only a few quarters. 
Remarkably, the number of small core-periphery pairs increases until the beginning of the global financial crisis, and then starts decreasing. 
Thus, the crisis event marks a tipping point in the tiered organisation of the interbank market. 
Additionally, the size of the main periphery decreases in time, indicating that main peripheral banks had an increasing tendency to move to small core-periphery pairs (or to become residual). 
Note also that, similarly to the static network case, for temporal networks the main core-periphery pair detected by the KM--ER algorithm 
mainly consists of Italian banks (Fig.~\ref{fig:italian_foreign_kmerb}). The second largest pair consists mainly of foreign banks, 
but this feature emerges only between 2002 and 2009 and for data aggregation periods that are not too short.

\begin{figure}
	\centering
	\includegraphics[width=\hsize]{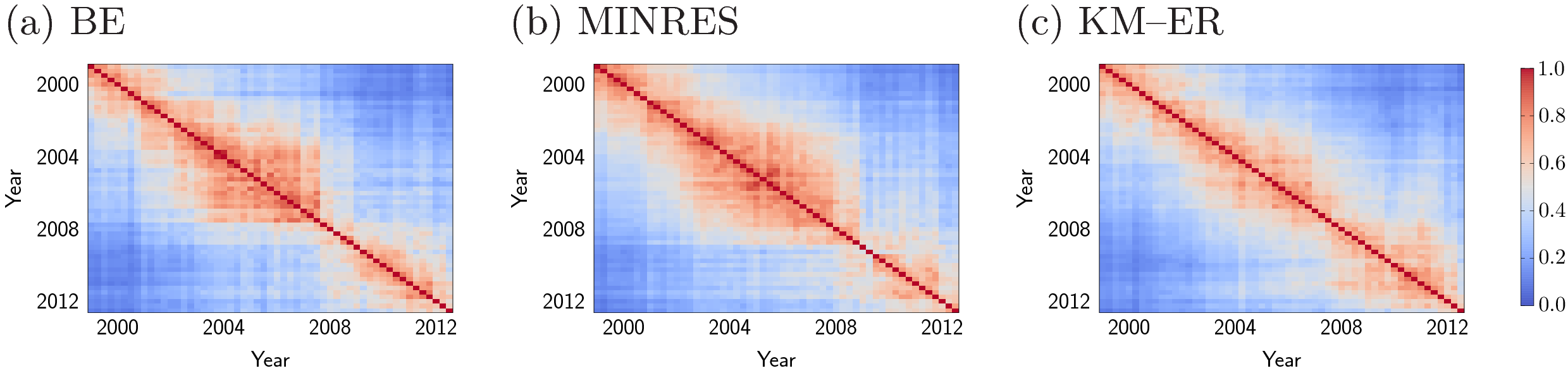}
\caption{Jaccard index for the composition of the main core (quarterly networks).}
\label{fig:jindex}
\end{figure}

To further inspect the effect of the crisis on eMID, we compute the Jaccard's coefficient for the bank composition of the main core at each pair of time points $t$ and $t'$, 
\ie, $|{{\mathcal{C}^{[1]}(t)}} \cap {{\mathcal{C}^{[1]}(t')}}| / |{{\mathcal{C}^{[1]}(t)}} \cup {{\mathcal{C}^{[1]}(t')}}|$, 
where ${\cal C}^{[1]}(t)$ is the set of core nodes of the main core-periphery pair at time $t$. 
The Jaccard index is large if the cores at times $t$ and $t'$ have many banks in common, and small otherwise. 
As shown in Figure~\ref{fig:jindex}, the main cores detected by all three algorithms experience an abrupt change around 2002 
(supposedly after the {\em dot.com} bubble burst) and especially in 2008, in the midst of the global financial crisis.

\begin{figure}
	\centering
	\includegraphics[width=\hsize]{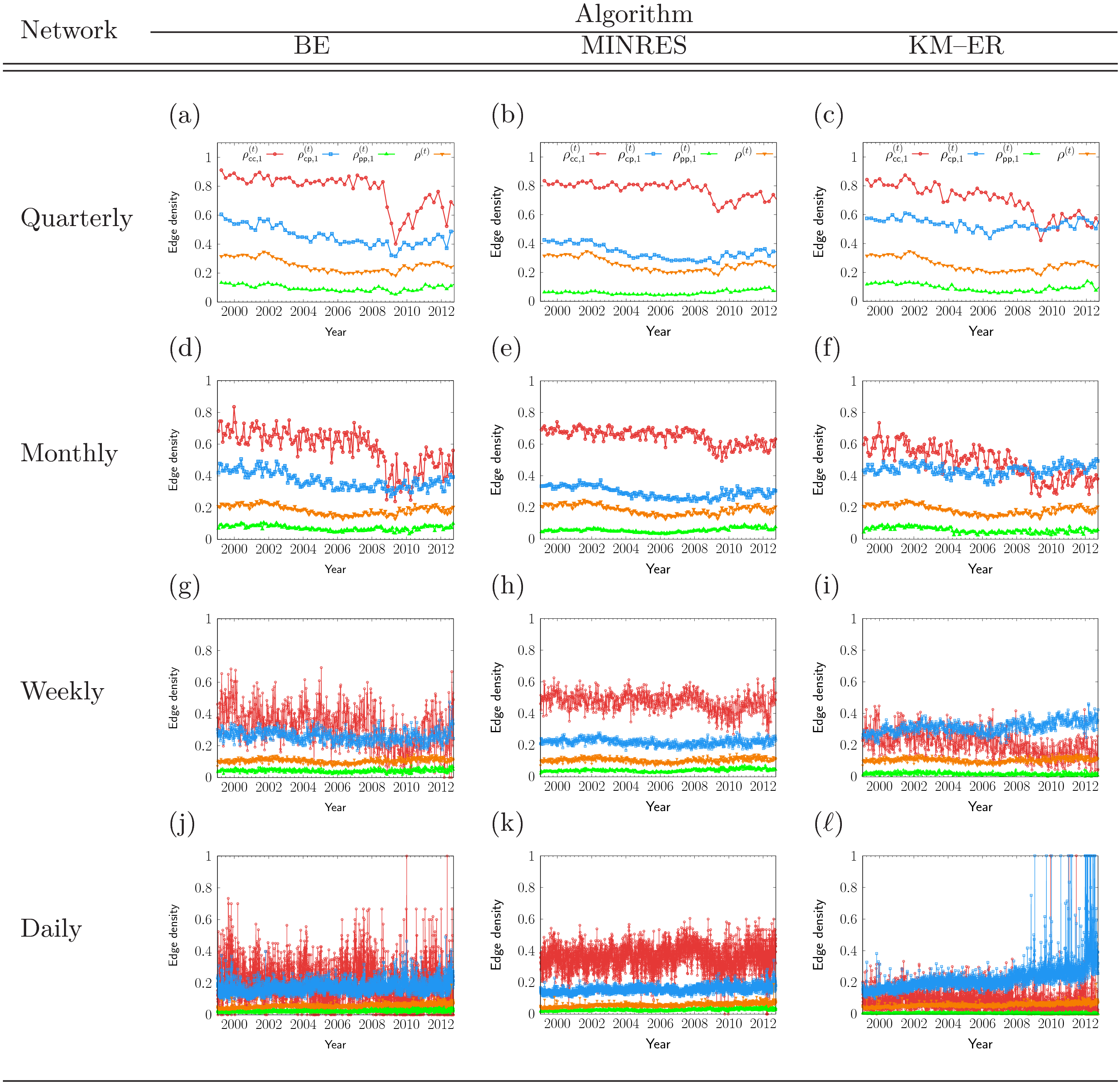}
\caption{Density of connections within and among different blocks, for the main core-periphery pair.}
\label{fig:density}
\end{figure}

We conclude our analysis by studying the internal structure of the main core-periphery pairs detected by the three algorithms.
To this end, we compute the fraction of edges (\ie, the edge density) within the core, that between the core and periphery, and that within the periphery in the main core-periphery pair. 
Figure~\ref{fig:density} shows the density the of different types of edges within the main core-periphery pairs. 
In most cases, we observe a marked decay in the density of intra-core edges after 2008, leading to a higher segregation of the network. 
Concerning the specifics of the algorithms, BE and MINRES always return a ``standard'' core-periphery structure, 
namely $\rhocc1>\rhocp1>\rhopp1$ (the less clear case of BE on daily networks will be discussed shortly). The KM--ER algorithm instead returns a more composite picture. 
For long aggregation periods (\ie, quarterly and monthly), a standard core-periphery structure is detected only {\em before} the crisis. 
Afterwards, we observe $\rhocp1>\rhocc1$: connections between core and periphery become more abundant than those within the core. 
This is a signature of bipartitivity, and again the structural transition happens during the crisis. For short aggregation periods (weekly and daily), bipartitivity is detected in the majority of cases, 
in accordance with the findings of \citet{Barucca2018}. 

Finally, focusing on daily networks (which are the noisiest due to the scarcity of connections), bipartite-like structures are detected by the BE, MINRES and KM--ER algorithms 
in approximately $27\%$, $2\%$ and $59\%$ of the cases, respectively.

\section{Discussion and Conclusions}

In this work we employed the KM--ER algorithm \citep{Kojaku2017} to characterise the internal organisation of the electronic market for interbank deposit eMID. 
As compared to other core-periphery detection algorithms, the KM--ER method stands out by identifying multiple core-periphery pairs 
allowing, for instance, to discern Italian and foreign banks in an unsupervised way. 
Note that although KM--ER is designed to detect core-periphery structures, the method has also been able to reveal the bipartite-like organisation of the market. 
This result is consistent with previous studies \citep{Barucca2016,Barucca2018}, in which bipartite-like structures in the eMID network were detected using 
stochastic block-modelling and its degree-corrected version \citep{Karrer2011}. In particular, using standard stochastic block-modelling \citet{Barucca2016,Barucca2018} found a single core-periphery pair 
which turned into a bipartite structure for short data aggregation periods, whereas, using degree-corrected stochastic block-modelling bipartitivity emerged also for longer periods. 
Here we found bipartite-like blocks in the networks by allowing multiple core-periphery pairs without discounting the effect of nodes' degree, 
and yet we detected bipartitivity for long data aggregation periods as well.  
Concerning the temporal evolution of the market, the KM--ER algorithm revealed a structural transition during the global financial crisis, 
both in terms of network segregation and organisation: the multiplicity of core-periphery pairs disappeared, and the main core-periphery structure was replaced by a bipartite-like structure. 

Clearly, the network patterns revealed by any algorithm-based analysis do depend on how the algorithm is designed. 
It would be useful to perform, in the future, an extensive comparison of the results obtained by both core-periphery and block-model based methods, 
in order to reveal the market features which are invariant from the observation lens used in the analysis.

\section*{Acknowledgements}
G.C. and G.C. acknowledge support from the EU projects DOLFINS (640772), CoeGSS (676547), Shakermaker (687941), and SoBigData (654024).
N.M. acknowledges the support provided through JST CREST Grant Number JPMJCR1304 and the JST ERATO Grant Number JPMJER1201, Japan. 
The funders had no role in study design, data collection and analysis, decision to publish, or preparation of the manuscript.

\section*{Declarations of Interest}
The authors report no conflicts of interest. The authors alone are responsible for the content and writing of the paper.

\bibliographystyle{apalike}

\begin{thebibliography}{}

\bibitem[Acemoglu et~al., 2015]{Acemoglu2015}
Acemoglu, D., Ozdaglar, A., and Tahbaz-Salehi, A. (2015).
\newblock Systemic risk and stability in financial networks.
\newblock {\em American Economic Review}, 105(2):564--608.

\bibitem[Acharya and Merrouche, 2013]{Acharya2013}
Acharya, V.~V. and Merrouche, O. (2013).
\newblock Precautionary hoarding of liquidity and interbank markets: Evidence
  from the subprime crisis.
\newblock {\em Review of Finance}, 17(1):107--160.

\bibitem[Adrian and Shin, 2010]{Adrian2010}
Adrian, T. and Shin, H.~S. (2010).
\newblock Liquidity and leverage.
\newblock {\em Journal of Financial Intermediation}, 19(3):418--437.

\bibitem[Allen and Gale, 2000]{Allen2000}
Allen, F. and Gale, D. (2000).
\newblock Financial contagion.
\newblock {\em Journal of Political Economy}, 108(1):1--33.

\bibitem[Allen et~al., 2014]{Allen2014}
Allen, F., Hryckiewicz, A., Kowalewski, O., and T\"umer-Alkan, G. (2014).
\newblock Transmission of financial shocks in loan and deposit markets: Role of
  interbank borrowing and market monitoring.
\newblock {\em Journal of Financial Stability}, 15:112--126.

\bibitem[Amini et~al., 2016]{Amini2016}
Amini, H., Cont, R., and Minca, A. (2016).
\newblock Resilience to contagion in financial networks.
\newblock {\em Mathematical Finance}, 26(2):329--365.

\bibitem[Anand et~al., 2015]{Anand2015}
Anand, K., Craig, B., and von Peter, G. (2015).
\newblock Filling in the blanks: Network structure and interbank contagion.
\newblock {\em Quantitative Finance}, 15(4):625--636.

\bibitem[Anand et~al., 2017]{Anand2017}
Anand, K., van Lelyveld, I., Banai, A., Christiano~Silva, T., Friedrich, S.,
  Garratt, R., Halaj, G., Hansen, I., Howell, B., Lee, H.,
  Mart\'{i}nez~Jaramillo, S., Molina-Borboa, J., Nobili, S., Rajan, S., Rubens
  Stancato~de Souza, S., Salakhova, D., and Silvestri, L. (2017).
\newblock The missing links: A global study on uncovering financial network
  structure from partial data.
\newblock {\em Journal of Financial Stability}, (in press).

\bibitem[Angelini et~al., 2011]{Angelini2011}
Angelini, P., Nobili, A., and Picillo, C. (2011).
\newblock The interbank market after august 2007: What has changed, and why?
\newblock {\em Journal of Money, Credit and Banking}, 43(5):923--958.

\bibitem[Bardoscia et~al., 2015]{Bardoscia2015}
Bardoscia, M., Battiston, S., Caccioli, F., and Caldarelli, G. (2015).
\newblock Debtrank: A microscopic foundation for shock propagation.
\newblock {\em PLoS ONE}, 10(6):e0130406.

\bibitem[Bardoscia et~al., 2017]{Bardoscia2017}
Bardoscia, M., Battiston, S., Caccioli, F., and Caldarelli, G. (2017).
\newblock Pathways towards instability in financial networks.
\newblock {\em Nature Communications}, 8:14416.

\bibitem[Barucca and Lillo, 2016]{Barucca2016}
Barucca, P. and Lillo, F. (2016).
\newblock Disentangling bipartite and core-periphery structure in financial
  networks.
\newblock {\em Chaos, Solitons and Fractals}, 88:244--253.

\bibitem[Barucca and Lillo, 2018]{Barucca2018}
Barucca, P. and Lillo, F. (2018).
\newblock The organization of the interbank network and how ecb unconventional
  measures affected the e-mid overnight market.
\newblock {\em Computational Management Science}, 15(1):33--53.

\bibitem[Battiston et~al., 2016]{Battiston2016complexity}
Battiston, S., Farmer, J.~D., Flache, A., Garlaschelli, D., Haldane, A.~G.,
  Heesterbeek, H., Hommes, C., Jaeger, C., May, R., and Scheffer, M. (2016).
\newblock Complexity theory and financial regulation.
\newblock {\em Science}, 351(6275):818--819.

\bibitem[Battiston et~al., 2012]{Battiston2012}
Battiston, S., Puliga, M., Kaushik, R., Tasca, P., and Caldarelli, G. (2012).
\newblock Debtrank: Too central to fail? financial networks, the fed and
  systemic risk.
\newblock {\em Scientific Reports}, 2:541.

\bibitem[Beaupain and Durr{\'e}, 2008]{Beaupain2008}
Beaupain, R. and Durr{\'e}, A. (2008).
\newblock The interday and intraday patterns of the overnight market: Evidence
  from an electronic platform.
\newblock ECB Working Paper Series 0988.

\bibitem[Bech and Atalay, 2010]{Bech2010}
Bech, M.~L. and Atalay, E. (2010).
\newblock The topology of the federal funds market.
\newblock {\em Physica A: Statistical Mechanics and its Applications},
  389(22):5223--5246.

\bibitem[Berrospide, 2013]{Berrospide2013}
Berrospide, J.~M. (2013).
\newblock Bank liquidity hoarding and the financial crisis: An empirical
  evaluation.
\newblock Finance and Economics Discussion Series~03, Board of Governors of the
  Federal Reserve System (U.S.).

\bibitem[Borgatti and Everett, 2000]{Borgatti2000}
Borgatti, S.~P. and Everett, M.~G. (2000).
\newblock Models of core/periphery structures.
\newblock {\em Social Networks}, 21(4):375--395.

\bibitem[Boss et~al., 2004]{Boss2004}
Boss, M., Elsinger, H., Summer, M., and Thurner, S. (2004).
\newblock Network topology of the interbank market.
\newblock {\em Quantitative Finance}, 4(6):677--684.

\bibitem[Boyd et~al., 2006]{Boyd2006}
Boyd, J.~P., Fitzgerald, W.~J., and Beck, R.~J. (2006).
\newblock Computing core/periphery structures and permutation tests for social
  relations data.
\newblock {\em Social Networks}, 28(2):165--178.

\bibitem[Boyd et~al., 2010]{Boyd2010}
Boyd, J.~P., Fitzgerald, W.~J., Mahutga, M.~C., and Smith, D.~A. (2010).
\newblock Computing continuous core/periphery structures for social relations
  data with minres/svd.
\newblock {\em Social Networks}, 32(2):125--137.

\bibitem[Brunnermeier, 2009]{Brunnermeier2009a}
Brunnermeier, M.~K. (2009).
\newblock Deciphering the liquidity and credit crunch 2007-2008.
\newblock {\em Journal of Economic Perspectives}, 23(1):77--100.

\bibitem[Caccioli et~al., 2014]{Caccioli2014}
Caccioli, F., Shrestha, M., Moore, C., and Farmer, J.~D. (2014).
\newblock Stability analysis of financial contagion due to overlapping
  portfolios.
\newblock {\em Journal of Banking and Finance}, 46:233--245.

\bibitem[Carre{\~n}o and Cifuentes, 2017]{Carreno2017}
Carre{\~n}o, J.~G. and Cifuentes, R. (2017).
\newblock Identifying complex core-periphery structures in the interbank
  market.
\newblock {\em Journal of Network Theory in Finance}, 3(4):49--75.

\bibitem[Chan-Lau et~al., 2009]{Lau2009}
Chan-Lau, J.~A., Espinosa, M., Giesecke, K., and Sol{\'e}, J.~A. (2009).
\newblock Assessing the systemic implications of financial linkages.
\newblock Technical report, IMF Global Financial Stability Report.

\bibitem[Cifuentes et~al., 2005]{Cifuentes2005}
Cifuentes, R., Ferrucci, G., and Shin, H.~S. (2005).
\newblock Liquidity risk and contagion.
\newblock {\em Journal of the European Economic Association}, 3(2/3):556--566.

\bibitem[Cimini and Serri, 2016]{Cimini2016}
Cimini, G. and Serri, M. (2016).
\newblock Entangling credit and funding shocks in interbank markets.
\newblock {\em PLoS ONE}, 11(8):e0161642.

\bibitem[Cimini et~al., 2015a]{Cimini2015estimating}
Cimini, G., Squartini, T., Gabrielli, A., and Garlaschelli, D. (2015a).
\newblock Estimating topological properties of weighted networks from limited
  information.
\newblock {\em Physical Review E}, 92:040802.

\bibitem[Cimini et~al., 2015b]{Cimini2015systemic}
Cimini, G., Squartini, T., Garlaschelli, D., and Gabrielli, A. (2015b).
\newblock Systemic risk analysis on reconstructed economic and financial
  networks.
\newblock {\em Scientific Reports}, 5:15758.

\bibitem[Cocco et~al., 2009]{Cocco2009}
Cocco, J.~F., Gomes, F.~J., and Martins, N.~C. (2009).
\newblock Lending relationships in the interbank market.
\newblock {\em Journal of Financial Intermediation}, 18(1):24--48.

\bibitem[Cont et~al., 2013]{Cont2013}
Cont, R., Moussa, A., and Santos, E.~B. (2013).
\newblock {\em Network structure and systemic risk in banking systems}, pages
  327--368.
\newblock Cambridge University Press.

\bibitem[Cont and Wagalath, 2016]{Cont2016}
Cont, R. and Wagalath, L. (2016).
\newblock Fire sales forensics: Measuring endogenous risk.
\newblock {\em Mathematical Finance}, 26(4):835--866.

\bibitem[Craig and Von~Peter, 2014]{Craig2014}
Craig, B. and Von~Peter, G. (2014).
\newblock Interbank tiering and money center banks.
\newblock {\em Journal of Financial Intermediation}, 23(3):322--347.

\bibitem[De~Masi et~al., 2006]{DeMasi2006}
De~Masi, G., Iori, G., and Caldarelli, G. (2006).
\newblock Fitness model for the italian interbank money market.
\newblock {\em Physical Review E}, 74:066112.

\bibitem[Di~Gangi et~al., 2015]{DiGiangi2015}
Di~Gangi, D., Lillo, F., and Pirino, D. (2015).
\newblock Assessing systemic risk due to fire sales spillover through maximum
  entropy network reconstruction.
\newblock available at \url{https://arxiv.org/abs/1509.00607}.

\bibitem[Diamond and Rajan, 2009]{Diamond2009}
Diamond, D.~W. and Rajan, R.~G. (2009).
\newblock Fear of fire sales and the credit freeze.
\newblock Working Paper 14925, National Bureau of Economic Research.

\bibitem[Elsinger et~al., 2006]{Elsinger2006}
Elsinger, H., Lehar, A., and Summer, M. (2006).
\newblock Risk assessment for banking systems.
\newblock {\em Management Science}, 52(9):1301--1314.

\bibitem[Finger et~al., 2013]{Finger2013}
Finger, K., Fricke, D., and Lux, T. (2013).
\newblock Network analysis of the e-mid overnight money market: The
  informational value of different aggregation levels for intrinsic dynamic
  processes.
\newblock {\em Computational Management Science}, 10(2):187--211.

\bibitem[Freixas et~al., 2000]{Freixas2000}
Freixas, X., Parigi, B.~M., and Rochet, J.-C. (2000).
\newblock Systemic risk, interbank relations, and liquidity provision by the
  central bank.
\newblock {\em Journal of Money, Credit and Banking}, 32(3):611--638.

\bibitem[Fricke and Lux, 2015]{Fricke2015}
Fricke, D. and Lux, T. (2015).
\newblock Core-periphery structure in the overnight money market: Evidence from
  the e-mid trading platform.
\newblock {\em Computational Economics}, 45(3):359--395.

\bibitem[Furfine, 2003]{Furfine2003}
Furfine, C.~H. (2003).
\newblock Interbank exposures: Quantifying the risk of contagion.
\newblock {\em Journal of Money, Credit and Banking}, 35(1):111--128.

\bibitem[Gai et~al., 2011]{Gai2011}
Gai, P., Haldane, A., and Kapadia, S. (2011).
\newblock Complexity, concentration and contagion.
\newblock {\em Journal of Monetary Economics}, 58(5):453--470.

\bibitem[Gai and Kapadia, 2010]{Gai2010}
Gai, P. and Kapadia, S. (2010).
\newblock Contagion in financial networks.
\newblock {\em Proceedings of the Royal Society of London A},
  466(2120):2401--2423.

\bibitem[Gale and Yorulmazer, 2013]{Gale2013}
Gale, D. and Yorulmazer, T. (2013).
\newblock Liquidity hoarding.
\newblock {\em Theoretical Economics}, 8(2):291--324.

\bibitem[Gandy and Veraart, 2016]{Gandy2015X}
Gandy, A. and Veraart, L.~A. (2016).
\newblock A bayesian methodology for systemic risk assessment in financial
  networks.
\newblock {\em Management Science (articles in advances)}, pages 1--20.

\bibitem[Georg, 2013]{Georg2013}
Georg, C.-P. (2013).
\newblock The effect of the interbank network structure on contagion and common
  shocks.
\newblock {\em Journal of Banking and Finance}, 37(7):2216--2228.

\bibitem[Glasserman and Young, 2015]{Glasserman2015}
Glasserman, P. and Young, H.~P. (2015).
\newblock How likely is contagion in financial networks?
\newblock {\em Journal of Banking and Finance}, 50:383--399.

\bibitem[Greenwood et~al., 2015]{Greenwood2015}
Greenwood, R., Landier, A., and Thesmar, D. (2015).
\newblock Vulnerable banks.
\newblock {\em Journal of Financial Economics}, 115(3):471--485.

\bibitem[Gualdi et~al., 2016]{Gualdi2016}
Gualdi, S., Cimini, G., Primicerio, K., Clemente, R.~D., and Challet, D.
  (2016).
\newblock Statistically validated network of portfolio overlaps and systemic
  risk.
\newblock {\em Scientific Reports}, 6:39467.

\bibitem[Haldane and May, 2011]{Haldane2011}
Haldane, A.~G. and May, R.~M. (2011).
\newblock Systemic risk in banking ecosystems.
\newblock {\em Nature}, 469:351--355.

\bibitem[H{\"u}ser, 2015]{Huser2015}
H{\"u}ser, A.-C. (2015).
\newblock Too interconnected to fail: A survey of the interbank networks
  literature.
\newblock {\em Journal of Network Theory In Finance}, 1(3):1--50.

\bibitem[Iori et~al., 2006]{Iori2006}
Iori, G., Jafarey, S., and Padilla, F.~G. (2006).
\newblock Systemic risk on the interbank market.
\newblock {\em Journal of Economic Behavior and Organization}, 61(4):525--542.

\bibitem[Iori et~al., 2015]{Iori2015}
Iori, G., Mantegna, R.~N., Marotta, L., MiccichÃ¨, S., Porter, J., and
  Tumminello, M. (2015).
\newblock Networked relationships in the e-mid interbank market: A trading
  model with memory.
\newblock {\em Journal of Economic Dynamics and Control}, 50:98--116.

\bibitem[Iori et~al., 2008]{Iori2008}
Iori, G., Masi, G.~D., Precup, O.~V., Gabbi, G., and Caldarelli, G. (2008).
\newblock A network analysis of the italian overnight money market.
\newblock {\em Journal of Economic Dynamics and Control}, 32(1):259--278.

\bibitem[Karrer and Newman, 2011]{Karrer2011}
Karrer, B. and Newman, M.~E.~J. (2011).
\newblock Stochastic blockmodels and community structure in networks.
\newblock {\em Physical Review E}, 83:016107.

\bibitem[Kaufman, 1994]{Kaufman1994}
Kaufman, G.~G. (1994).
\newblock Bank contagion: A review of the theory and evidence.
\newblock {\em Journal of Financial Services Research}, 8(2):123--150.

\bibitem[Kojaku and Masuda, 2017]{Kojaku2017}
Kojaku, S. and Masuda, N. (2017).
\newblock Finding multiple core-periphery pairs in networks.
\newblock {\em Physical Review E}, 96:052313.

\bibitem[Krause and Giansante, 2012]{Krause2012}
Krause, A. and Giansante, S. (2012).
\newblock Interbank lending and the spread of bank failures: A network model of
  systemic risk.
\newblock {\em Journal of Economic Behavior and Organization}, 83(3):583--608.

\bibitem[Langfield et~al., 2014]{Langfield2014}
Langfield, S., Liu, Z., and Ota, T. (2014).
\newblock Mapping the uk interbank system.
\newblock {\em Journal of Banking and Finance}, 45:288--303.

\bibitem[Lee, 2013]{Lee2013}
Lee, S.~H. (2013).
\newblock Systemic liquidity shortages and interbank network structures.
\newblock {\em Journal of Financial Stability}, 9(1):1--12.

\bibitem[Le{\'o}n and Berndsen, 2014]{Leon2014}
Le{\'o}n, C. and Berndsen, R.~J. (2014).
\newblock Rethinking financial stability: Challenges arising from financial
  networks' modular scale-free architecture.
\newblock {\em Journal of Financial Stability}, 15:241--256.

\bibitem[Lip, 2011]{Lip2011}
Lip, S.~Z. (2011).
\newblock A fast algorithm for the discrete core/periphery bipartitioning
  problem.
\newblock available at \url{https://arxiv.org/abs/1102.5511}.

\bibitem[Martinez-Jaramillo et~al., 2014]{MartinezJaramillo2014}
Martinez-Jaramillo, S., Alexandrova-Kabadjova, B., Bravo-Benitez, B., and
  Sol\'orzano-Margain, J.~P. (2014).
\newblock An empirical study of the mexican banking system'2 network and its
  implications for systemic risk.
\newblock {\em Journal of Economic Dynamics and Control}, 40:242--265.

\bibitem[Nier et~al., 2007]{Nier2007}
Nier, E.~W., Yang, J., Yorulmazer, T., and Alentorn, A. (2007).
\newblock Network models and financial stability.
\newblock {\em Journal of Economic Dynamics and Control}, 31(6):2033--2060.

\bibitem[Rochet and Tirole, 1996]{Rochet1996}
Rochet, J.-C. and Tirole, J. (1996).
\newblock Interbank lending and systemic risk.
\newblock {\em Journal of Money, Credit and Banking}, 28(4):733--762.

\bibitem[Serri et~al., 2017]{Serri2017}
Serri, M., Caldarelli, G., and Cimini, G. (2017).
\newblock How the interbank market becomes systemically dangerous: an
  agent-based network model of financial distress propagation.
\newblock {\em Journal of Network Theory in Finance}, 3(1):1--15.

\bibitem[{\v{S}}id{\'{a}}k, 1967]{Sidak1967}
{\v{S}}id{\'{a}}k, Z. (1967).
\newblock Rectangular confidence regions for the means of multivariate normal
  distributions.
\newblock {\em Journal of the American Statistical Association},
  62(318):626--633.

\bibitem[Silva et~al., 2016]{Silva2016}
Silva, T.~C., de~Souza, S. R.~S., and Tabak, B.~M. (2016).
\newblock Network structure analysis of the brazilian interbank market.
\newblock {\em Emerging Markets Review}, 26:130--152.

\bibitem[Soram\"aki et~al., 2007]{Soramaki2007}
Soram\"aki, K., Bech, M.~L., Arnold, J., Glass, R.~J., and Beyeler, W.~E.
  (2007).
\newblock The topology of interbank payment flows.
\newblock {\em Physica A: Statistical Mechanics and its Applications},
  379(1):317--333.

\bibitem[Squartini et~al., 2017]{Squartini2017ECAPM}
Squartini, T., Almog, A., Caldarelli, G., van Lelyveld, I., Garlaschelli, D.,
  and Cimini, G. (2017).
\newblock Enhanced capital-asset pricing model for the reconstruction of
  bipartite financial networks.
\newblock {\em Physical Review E}, 96:032315.

\bibitem[Veld and van Lelyveld, 2014]{Veld2014}
Veld, D. i.~t. and van Lelyveld, I. (2014).
\newblock Finding the core: Network structure in interbank markets.
\newblock {\em Journal of Banking and Finance}, 49:27--40.

\bibitem[Verma et~al., 2016]{Verma2016}
Verma, T., Russmann, F., Ara{\'u}jo, N., Nagler, J., and Herrmann, H. (2016).
\newblock Emergence of core-peripheries in networks.
\newblock {\em Nature Communications}, 7:10441.

\end{thebibliography}


\newpage

\appendix

\renewcommand{\thefigure}{A.\arabic{figure}} 

\begin{figure}
\centering
\scalebox{0.9}{	\includegraphics[width=\hsize]{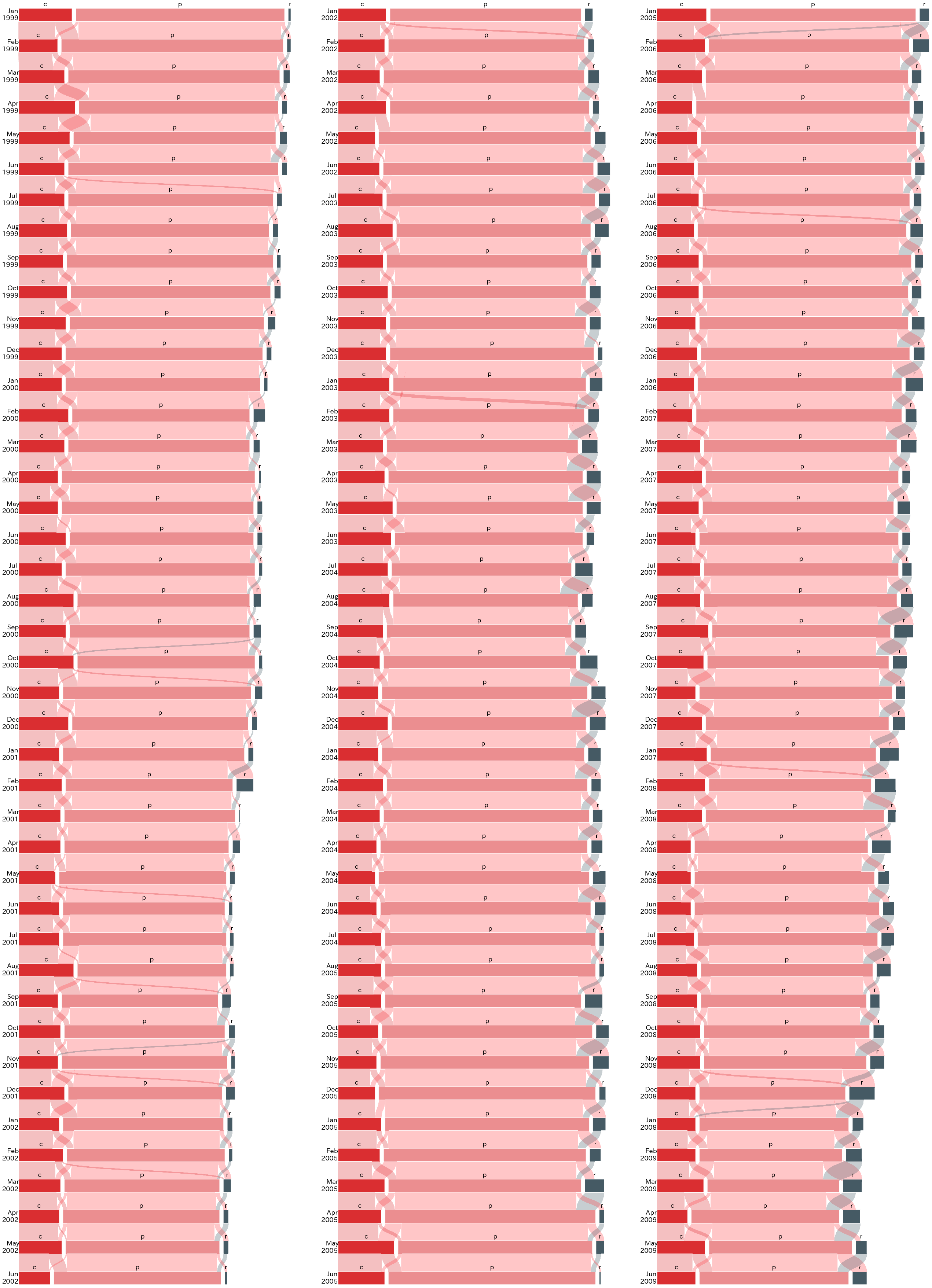} }
\caption{Core-periphery structure in monthly networks detected by the BE algorithm. See the caption of Fig.~\ref{fig:allu_quarterly} for legends.}
\label{fig:allu_beb_monthly}
\end{figure}

\clearpage 

\begin{figure}
\centering
\scalebox{0.9}{	\includegraphics[width=\hsize]{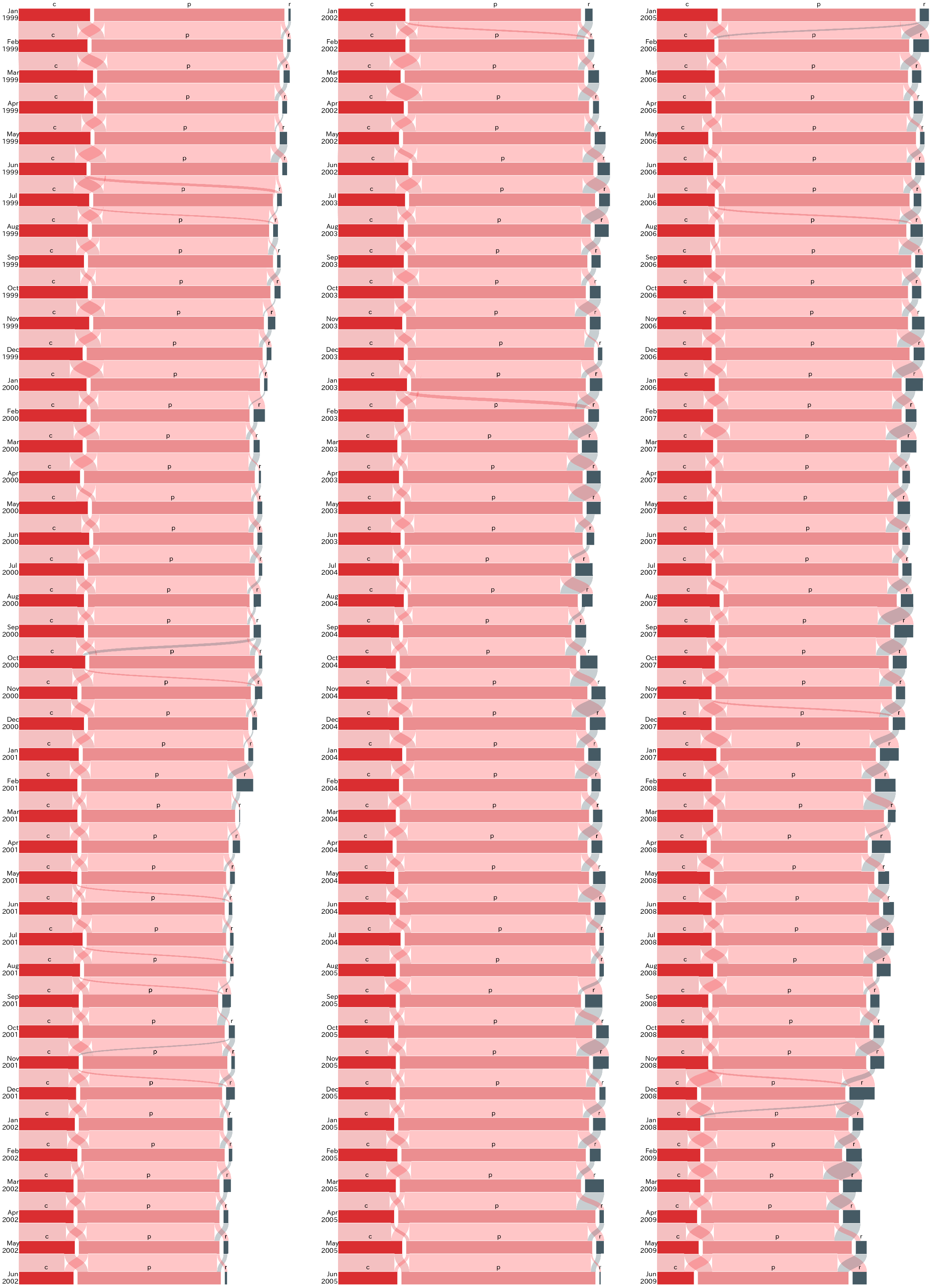} }
\caption{Core-periphery structure in monthly networks detected by the MINRES algorithm. See the caption of Fig.~\ref{fig:allu_quarterly} for legends.}
\label{fig:allu_minresb_monthly}
\end{figure}

\clearpage 

\begin{figure}
\centering
\scalebox{0.9}{	\includegraphics[width=\hsize]{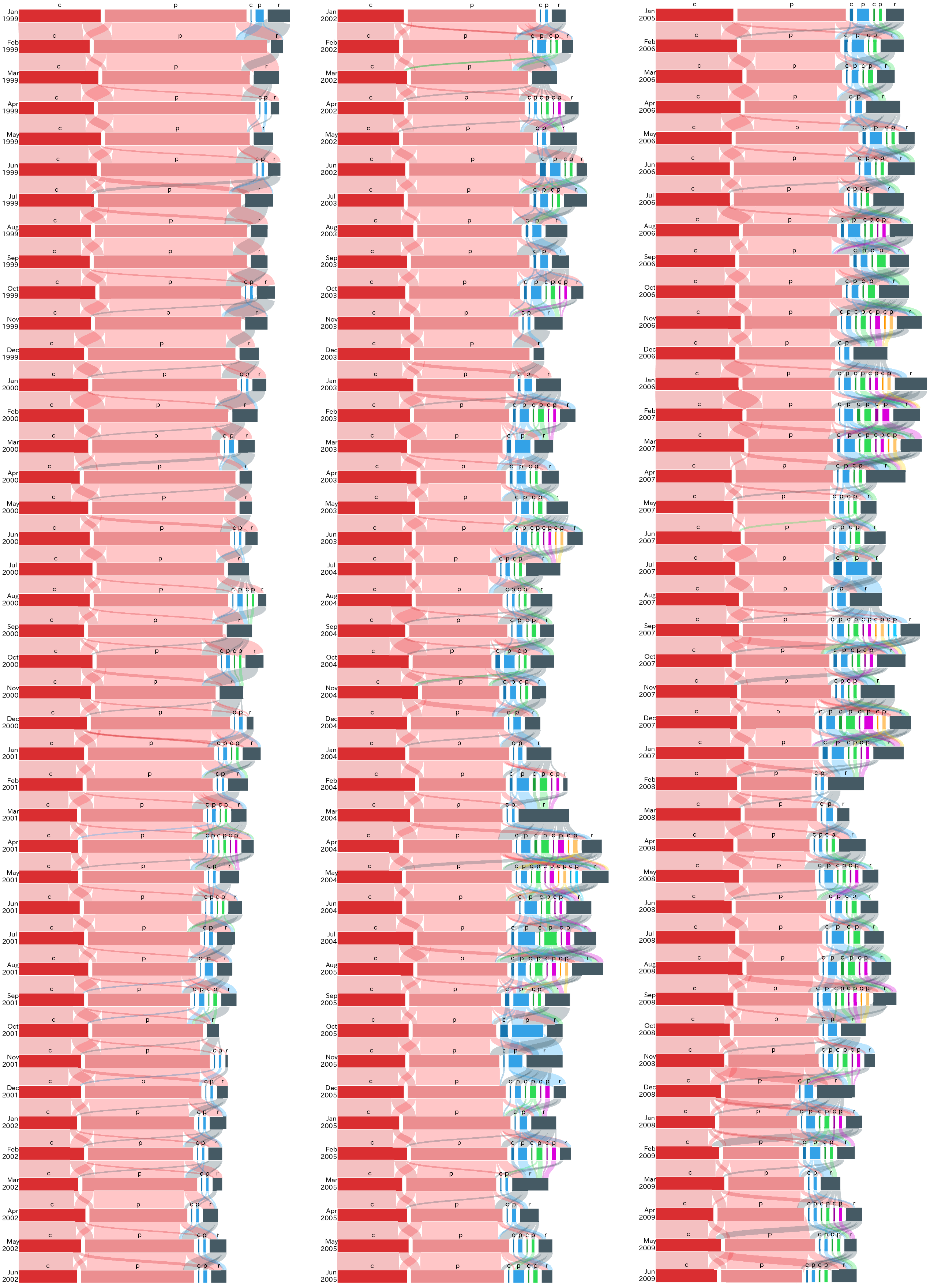} }
\caption{Core-periphery structure in monthly networks detected by the KM--ER algorithm. See the caption of Fig.~\ref{fig:allu_quarterly} for legends.}
\label{fig:allu_kmerb_monthly}
\end{figure}

\end{document}